\documentclass[conference]{IEEEtran}
\IEEEoverridecommandlockouts
\usepackage{url}
\usepackage{enumitem}
\usepackage{cite}
\usepackage{amsmath,amssymb,amsfonts}
\usepackage{algorithmic}
\usepackage{graphicx}
\usepackage{textcomp}
\usepackage{xcolor}
\usepackage{float}
\usepackage[utf8]{inputenc}
\def\BibTeX{{\rm B\kern-.05em{\sc i\kern-.025em b}\kern-.08em
    T\kern-.1667em\lower.7ex\hbox{E}\kern-.125emX}}
\begin{document}

\title{A Taxonomy of Approaches for Integrating Attack Awareness in Applications} 

\author{\IEEEauthorblockN{Tolga Ünlü,
Lynsay A. Shepherd, Natalie Coull, Colin McLean}
\IEEEauthorblockA{Division of Cyber Security, School of Design and Informatics\\
Abertay University\\
Dundee, United Kingdom\\
Email: 1205365@abertay.ac.uk,
lynsay.shepherd@abertay.ac.uk,
n.coull@abertay.ac.uk,
c.mclean@abertay.ac.uk}}

\maketitle

\begin{abstract}
Software applications are subject to an increasing number of attacks, resulting in data breaches and financial damage.  Many solutions have been considered to help mitigate these attacks, such as the integration of attack-awareness techniques.  In this paper, we propose a taxonomy illustrating how existing attack awareness techniques can be integrated into applications.  This work provides a guide for security researchers and developers, aiding them when choosing the approach which best fits the needs of their application.
\end{abstract}

\smallskip 


\section{Introduction} \label{introduction}
To build applications that can cope with a continuous and dynamic threat landscape, it is vital to empower developers with methodologies and tools for the development of effective defences \cite{green2016}. It is also essential that these methodologies and tools can handle the fast-paced nature of modern development practices, such as Continuous Integration (CI) and Continuous Delivery (CD).  These practices are often applied within a DevOps environment in which collaboration between software development and IT operations, including security, is encouraged\cite{ebert2016}. If the current tools and methodologies remain unchanged, security will be perceived as slowing down the development process, making it less appreciated by developers and thus leading to insecure software applications.  

Organizations which utilize a DevOps culture typically have an overarching view regarding what is deployed in their environment.  However, this usually focuses on the performance of applications, or errors caused within applications. To gain similar insights on security-related events, it is necessary to integrate attack awareness into an application. Otherwise, ongoing attacks may be missed, and there is the potential for a compromise or data breach to go unnoticed. Data breaches can have a substantial financial and reputational impact on an organization- according to IBM Security\cite{ibmsecurity2019}, a breach which takes 200 days to resolve can cost a company \$4.56 million. 

This paper discusses a taxonomy of approaches regarding how to integrate attack awareness into applications, providing a guide for developers and security researchers.  The remainder of the paper is structured as follows: Section \ref{problemstatement} describes the need for attack aware applications.  Current approaches are described in Section \ref{relatedwork}.  Section \ref{analysis} provides an analysis of these solutions before conclusions are drawn in Section \ref{conclusion}.
 
\section{Problem Statement} \label{problemstatement}
Many existing applications cannot provide real-time intelligence regarding their security state \cite{Watson2011}. This lack of information means developers do not know whether an attack has taken place, or which parts of an application attracts the most attention from those with malicious intentions.  Integrating attack awareness can address this knowledge gap, providing developers with actionable insights based on the context of the application. 

Integration is generally performed by the developer or by utilizing an agent which makes an application attack-aware at runtime. While a fully automated approach is desirable, it is only feasible for application components developed using established practices and technologies \cite{zhu2016}.  Additionally, off-the-shelf protection methods implemented by agents are limited to specific attack classes \cite{noderasp20182}\cite{hawkins2017} and known attacks \cite{openrasp20182}. A taxonomy of approaches can guide researchers and developers in their choice of an appropriate integration method, taking into consideration the usability of the integration method.

\section{Related Work} \label{relatedwork}
Approaching intrusion detection from within an application has been hypothesized by Sielken \cite{Sielken1999} to detect attackers that abuse target applications without exposing anomalous behavior, and by avoiding the use of detectable attack patterns. Detection of these attacks can be achieved by utilizing contextual information such as the methods invoked while using a certain application feature. There are two approaches proposed to collect contextual application information.  The first approach regularly scans internal values of the target application, such as the authorization level of an user, and documents any changes of these values. The second approach embeds code triggers directly into the target applications' code. These can be conditional checks such as whether a predefined option that has been selected by a user has been modified.  Work by Kerschbaum et al. \cite{kerschbaum2002} provides an example of code triggers usage to detect network-based attacks within the OpenBSD kernel, and application-specific attacks from within Sendmail. The AppSensor framework \cite{Watson2011}, and BlackWatch \cite{Hall2019} are further examples which apply the code trigger variant.

The concept of reference monitoring, as defined by Anderson \cite{Anderson1972}, enforces an access policy upon user programs in execution that access references such as other programs, data or peripherals. Originally, reference monitors were implemented at operating system (OS) level to enforce a security policy on applications interacting with the OS kernel. However, there are two further approaches- the reference monitor is either embedded in an interpreter that runs the target application or it is embedded in the target application directly \cite{Erlingsson2003}.

Self-protecting software systems are another closely related field of research, focusing on systems which autonomously defend themselves against attacks, as envisioned by Kephart and Chess\cite{Kephart2003}. Furthermore, these systems can also anticipate security risks and mitigate them by proactively enabling countermeasures. In the nine dimensional taxonomy of self-protecting software systems by Yuan and Malek \cite{Yuan2012}, the authors dedicate one dimension for research which focuses on achieving self-protection at a specific phase within the Software Development Lifecycle (SDLC). These phases are classified as design, development, testing and implementation, but are simplified in Yuan and Malek's \cite{Yuan2012} work as self-protection achieved at development-time and runtime. 

\section{Analysis} \label{analysis}
\begin{figure*}
    \center
    \includegraphics[height=4.5cm]{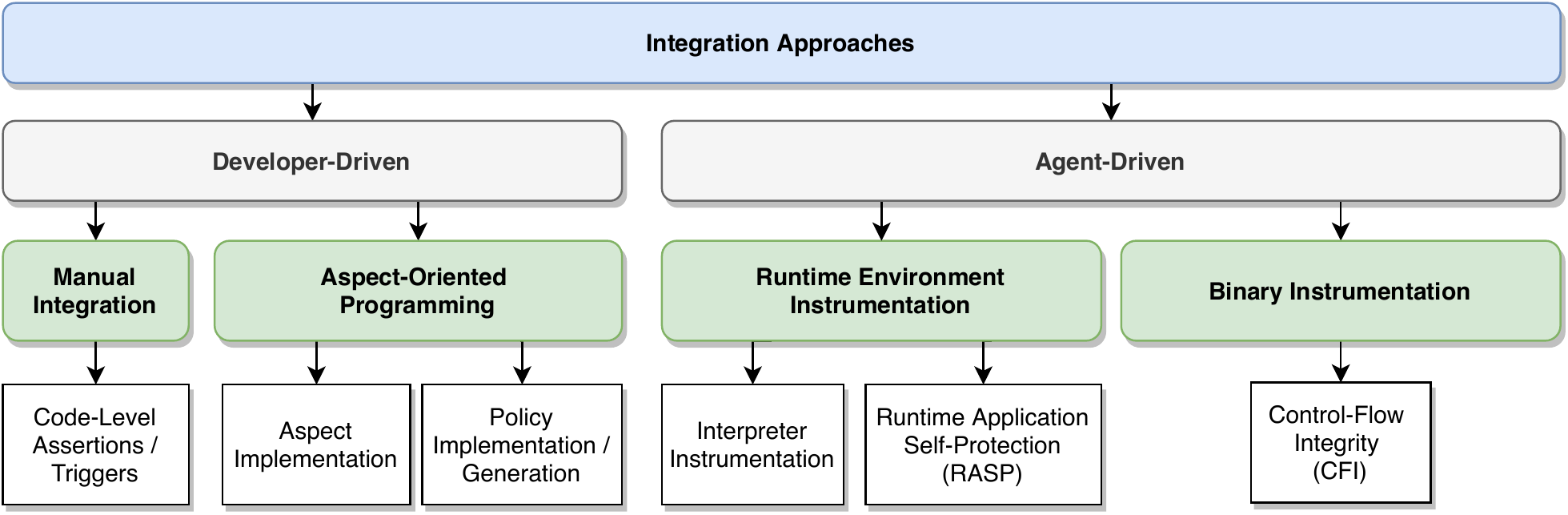}
    \caption{Developer and agent-driven approaches to integrate attack awareness into an application}
    \label{fig:intfig}
\end{figure*}

This section provides an analysis of previous research and existing solutions, summarized in Figure \ref{fig:intfig}, creating a taxonomy of approaches for integrating attack-awareness in applications. Based on the research in section \ref{relatedwork}, the integration is performed by a developer or by an agent.  In the proposed taxonomy, it is assumed there is access to the source code of the target application required for the developer-driven approaches.

\textbf{Developer-Driven: }Developers of an application can integrate attack awareness by implementing security controls in the target application, or by configuring existing security controls within the target application. The main body of work in developer-driven attack awareness is conducted manually, and more detail is provided in subsections \ref{manualintegration} and \ref{aspectorientedprogramming}.

\textbf{Agent-Driven: }An agent (a software component or a software application which autonomously acts on behalf of its user) can also be designed to integrate attack awareness. Though the focus is on autonomous integration, the agent can provide an interface to enable manual configuration, generating security controls at runtime. An agent that is part of the interpreter can provide attack awareness through the runtime environment and thus to any application running in this environment. Subsection \ref{runtimeinstrumentation} describes agents integrating attack awareness through instrumentation. 

\subsection{Manual Integration} \label{manualintegration}
Solutions like the AppSensor framework \cite{Watson2011} and BlackWatch \cite{Hall2019} rely on the hypothesis that an attacker or malicious behavior can be detected by knowing the `normal' behavior of the target application, and by being able to monitor deviations from such behavior. Application developers are therefore ideal candidates to apply this integration approach as they have specified, designed and implemented the application, and know where to strategically place security controls.

While this approach is effective against attackers targeting the business logic of an application, it is less so when it comes to attacks such as injection attacks, which require developers to have expertise in their execution. This, however, is often not the case for security expertise among developers in general as argued by Wurster and Oorschot \cite{Wurster2008} - developers may lack security expertise.

It is the developers responsibility to put the security controls in place, however there is a chance that the placement is neglected due to a lack of priority or time spent on non-security tasks, leading to areas of the application which are not covered. In the worst case, this approach will not be accepted by the developers as it can be seen as an additional effort which needs to be completed on top of others tasks \cite{Hall2019}\cite{zhu2016}. Due to this issue, integration approaches should utilize proven development techniques e.g. Kim et al. \cite{kim2012} utilizes the dependency injection technique to make the integration of attack awareness scalable and reusable. This technique is part of modern application frameworks and while it is familiar to many developers, it can also address the issue of accidentally omitting areas of the application. From the perspective of framework developers, it is suggested to build the security controls into the framework.  According to an empirical study by Peguero et al. \cite{peguero2018}, applications using built-in security controls derived from a framework are less susceptible to vulnerabilities than applications where developers were responsible for implementing security controls. 

\subsection{Aspect-Oriented Programming} \label{aspectorientedprogramming}
Aspect-Oriented Programming (AOP) is a technique to separate cross-cutting concerns, such as logging and monitoring, which are used throughout an application but which do not represent any business functionality \cite{Kiczales1997}. These cross-cutting concerns, also referred to as aspects, are weaved in to the target functionality of the application by intercepting function or method calls at runtime.

Security controls can also be defined as aspects as they are not tied to the target application's business functionality. Serme et al.\cite{Serme2014} present an AOP approach on integrating input validation controls, whereas Phung et al. \cite{Phung2009} demonstrate aspectized security policies which prevent malicious behaviour in the target application.  

AOP reduces the manual effort required when initially implementing the security policies that are processed by aspects. An aspect can be implemented as code in the programming language of the target application \cite{Serme2014}\cite{Phung2009} but it can also be implemented in a modeling language such as Unified Modeling Language (UML)\cite{zhu2009}, or in a Program Query Language (PQL) \cite{livshits2006}. The various implementation methods can address the usability requirements of the different roles involved with the development of an application. A developer may prefer to implement an aspect as code whereas an architect may rather implement an aspect in a modeling language. Regardless of the role, the implementation method of choice should be usable for its target audience as highlighted by Viega et al.\cite{Viega2001}, to prevent developer-induced errors and to reduce the required expertise to implement security controls.  

Instead of manually implementing an aspect from scratch, it can also be generated using, for example, the output of a static code analysis tool \cite{simic2013}. While this method seems to automate the integration process at first, in reality it shifts the manual effort to the usage of a tool or application which generates the output required for the aspect generation. Whether the integration can be fully automated depends on the degree of automation supported by the tool or application in question.  

Iraqi and Bakkali \cite{iraqi2019} developed a framework that can learn to detect outlier behavior in method invocations without supervision. The method invocation features required for the learning process are implemented as a feature extractor aspect, which demonstrates another use case for integrating attack awareness with AOP.  

\subsection{Runtime Environment and Binary Instrumentation} \label{runtimeinstrumentation}
The approaches analysed in the previous sections are effective when integrating security controls into a single application at one time. However, integrating security controls from within the runtime environment can affect any application running in the environment. Existing solutions are based on a modified interpreter as in ZenIDS \cite{hawkins2017} or Node RASP \cite{noderasp2018}. Another variant is also known as a Runtime Application Self-Protection (RASP) agent \cite{gartner2020}, a library which can be loaded by an interpreter at runtime e.g. OpenRASP \cite{openrasp2018} or Sqreen \cite{jeanbaptiste2019}.

In the ideal case, RASP agents can be deployed in a plug and play manner, requiring only an initial configuration as Haupert et al. \cite{haupert2018} describes regarding the deployment of Promon SHIELD RASP \cite{promon2020}. In cases where an agent does not require any configuration or a learning phase, attacks are detected using techniques that, e.g., combine taint-tracking with lexical analysis \cite{noderasp20182} or that monitor common input sinks and output sources for known malicious behavior and signatures \cite{openrasp20182}. Depending on the RASP implementation, the aforementioned techniques may only cover a limited set of sinks and sources, and the detection of known attacks relies on the completeness of the malicious behavior or signature database available. As RASP solutions are platform specific by design, it can also be the case that specific technologies used within a target application, or the entire application, might not be able to benefit from this integration approach.

While an instrumented runtime environment may be suitable for applications written in interpreted languages, those written in compiled languages require individual instrumentation.  With binary instrumentation, security controls can be integrated into an applications binary code to guarantee control-flow integrity (CFI) at runtime \cite{abadi2009}. The effectiveness of CFI is limited to attacks attempting to hijack the control flow of an application, such as buffer overflows.  Although it is feasible to integrate CFI in closed source applications, many available CFI implementations require access to source code\cite{goktas2014}, making it impractical for legacy or third-party applications.

\subsection{Summary}
The previous subsections have shown that all attack awareness approaches require some form of manual interaction, either at the setup or configuration phase. This is particularly pertinent for the detection of attacks against the business logic of an application, as these are unique for every application. The runtime environment and binary instrumentation approaches are no exception and also require manual intervention by the developers in the form of placing function calls which can track authenticated users\cite{hawkins2017} or custom events\cite{sqreen2020}. As with AOP approaches, it is then up to the agent implementation to provide a method of specifying policies, acting on the tracked events. This will influence how usable that method is in the context of the application.

Although the agent-driven approach has the most potential for automation, there are other aspects which make manual integration worthwhile, e.g. the insertion of code triggers can be completed in a few lines of non-fragmented code\cite{kerschbaum2002}. Compared to instrumented monitoring, code triggers are only called when they lie in the execution path of an attack and thus cause no performance overhead during non-malicious interactions. The code triggers are an inherent part of the application, providing constant attack-awareness in any environment in which the application is deployed.

The programming language can also be relevant when choosing AOP for attack awareness integration. To implement AOP in JavaScript, one can overwrite built-in functions, which makes self-protection possible \cite{Phung2009}. In contrast, PHP does not support the same functionality for built-in functions- a language extension must be installed first. Extension usage might not be feasible in certain environments due to restrictions in what can be deployed, or the lack of control over the environment. This would also be an exclusion criteria for choosing a RASP agent in the previously described environment. 

The approaches thus far focused on using attack awareness for application protection. However, the same concept can be utilized for security testing e.g. the insights provided by an agent could be used for the generation of security tests. Since the tester is an established role in modern development environments, enhancing the capabilities of such individuals could be a method to improve the adoption of attack awareness. 

\section{Conclusion and Future Work} \label{conclusion}
Work presented in this paper explored different methods of including attack awareness within applications. The taxonomy highlights that while there are a number of potential solutions, they do not necessarily meet the needs of developers. 
Research currently being undertaken by Ünlü \cite{Unlu2019} seeks to enhance the field of attack awareness and software security. As part of this process, the work will involve the investigation of techniques which aim to enhance the process by which developers include attack awareness within applications. By achieving an enhancement in the adoption of attack awareness techniques, the proposed research will contribute to bringing application security closer to developers by creating the necessary tools for them.  Such tools will be tailored to the needs of developers, empowering them to build secure applications.

\bibliographystyle{plain}
\bibliography{bibliography.bib}

\end{document}